\documentclass[journal]{IEEEtran}
\ifCLASSINFOpdf
\else
   \usepackage[dvips]{graphicx}
\fi
\usepackage{url}

\usepackage{booktabs}
\usepackage{mathrsfs}
\usepackage{array}
\usepackage{graphicx}
\usepackage{latexsym}
\usepackage{amsmath}
\usepackage{amsthm}
\usepackage{color}
\usepackage{xcolor}
\usepackage{amsfonts}
\usepackage{subfigure}
\usepackage{dsfont}
\usepackage{epstopdf}
\usepackage{threeparttable}
\usepackage{multirow}
\usepackage{enumerate}
\usepackage{epsfig}
\usepackage{bm}
\usepackage{cite}
\usepackage{algorithmic}
\usepackage{textcomp}
\usepackage{graphicx}
\usepackage{latexsym}
\usepackage{amsmath}
\usepackage{amsthm}
\usepackage{makecell,rotating,multirow,diagbox}
\usepackage{amsfonts}
\usepackage{subfigure}
\usepackage{dsfont}
\usepackage{epstopdf}
\usepackage{threeparttable}
\usepackage{multirow}
\usepackage{enumerate}
\usepackage{epsfig}
\usepackage{bm}
\usepackage{cite}
\usepackage{algorithmic}
\usepackage{textcomp}
\usepackage{xcolor}
\usepackage{amsmath}
\usepackage{cite}

\usepackage{tabu}
\usepackage{makecell}

\ifCLASSINFOpdf
\else
\fi
\begin{document}
%
\title{Fusion Learning for 1-Bit CS-based Superimposed CSI Feedback with Bi-Directional\\Channel Reciprocity}
%
%
%
\author{{Chaojin Qing,~\IEEEmembership{Member,~IEEE}, Qing Ye, Wenhui Liu, and Jiafan Wang}

\thanks{This work is supported in part by the Sichuan Science and Technology Program (Grant No. 2021JDRC0003), the Major Special Funds of Science and Technology of Sichuan Science and Technology Plan Project (Grant No. 19ZDZX0016 /2019YFG0395), the Demonstration Project of Chengdu Major Science and Technology Application (Grant No. 2020-YF09- 00048-SN), the Key Scientific Research Fund of Xihua University (Grant No. Z1120941), and the Special Funds of Industry Development of Sichuan Province (Grant No. zyf-2018-056).}
\thanks{C. Qing, Q. Ye and W. Liu are with School of Electrical Engineering and Electronic Information, Xihua University, Chengdu, China (E-mail: qingchj@mail.xhu.edu.cn).}

\thanks{Jiafan Wang is with Synopsys Inc., 2025 NE Cornelius Pass Rd, Hillsboro, OR 97124, USA (E-mail: jifanw@gmail.com).}
}

%
%

 \markboth{IEEE XXXX XXXX,~Vol.~XX, No.~XX, XXX~2022}%
 {Shell \MakeLowercase{\textit{et al.}}: Bare Demo of IEEEtran.cls for IEEE Journals}

%



\maketitle

\begin{abstract}
Due to the discarding of downlink channel state information (CSI) amplitude and the employing of iteration reconstruction algorithms, 1-bit compressed sensing (CS)-based superimposed CSI feedback is challenged by low recovery accuracy and large processing delay.
To overcome these drawbacks, this letter proposes a fusion learning scheme by exploiting the bi-directional channel reciprocity. Specifically, a simplified version of the conventional downlink CSI reconstruction is utilized to extract the initial feature of downlink CSI, and a single hidden layer-based amplitude-learning network (AMPL-NET) is designed to learn the auxiliary feature of the downlink CSI amplitude.
Then, based on the extracted and learned amplitude features, a simple but effective amplitude-fusion network (AMPF-NET) is developed to perform the amplitude fusion of downlink CSI and thus improves the reconstruction accuracy for 1-bit CS-based superimposed CSI feedback
while reducing the processing delay.
Simulation results show the effectiveness of the proposed feedback scheme and the robustness against parameter variations.
\end{abstract}

\begin{IEEEkeywords}
Channel state information (CSI), 1-bit compressed sensing (CS), superimposed CSI feedback, fusion learning, bi-directional reciprocal channel characteristics.
\end{IEEEkeywords}

\IEEEpeerreviewmaketitle

\section{Introduction}

\IEEEPARstart{A}{s} one of the key technologies of the fifth generation (5G) and beyond wireless communication systems, massive multiple-input multiple-output (mMIMO) systems have shown great prospects in providing high spectrum and energy efficiency \cite{a1}. However, to take full advantage of the system, mMIMO transmitters must rely on sufficiently accurate channel state information (CSI), which means that the base station (BS) needs to acquire the downlink CSI accurately and timely \cite{d1}.

In frequency division duplex (FDD) systems, the downlink CSI usually needs to be fed back to the BS. However, numerous BS antennas consume huge feedback overhead\cite{d2}.
To improve feedback efficiency, different neural network (NN)-based CSI feedbacks have been aroused in recent years \cite{d1,d2,d3}. Yet the uplink bandwidth resource is still heavily occupied due to the mMIMO scenarios.
This dilemma is overcome by employing superimposed CSI feedback \cite{c1,x1,c2}, while introducing the superimposed interference between downlink CSI and uplink user data sequences (UL-US).
To alleviate the superimposed interference, 1-bit compressed sensing (CS)-based superimposed CSI feedback is proposed in \cite{y1} by transforming the estimation problem into a detection problem. Thus, the superimposed interference is effectively reduced, especially for the detection of UL-US.

Nevertheless, the reconstruction of downlink CSI in 1-bit CS-based superimposed CSI feedback is still facing huge challenges. 1) The amplitude information is discarded, resulting in the reconstruction accuracy of downlink CSI is unsatisfactory. 2) The commonly used iterative reconstruction for 1-bit CS causes excessive processing delay.
These issues motivate the following considerations. 1) The discarded amplitude information in 1-bit CS needs to be retrieved as much as possible without increasing the additional uplink bandwidth. This promotes us to exploit the multiple modals of amplitude features. 2) To reduce the processing delay, the structures of the developed NNs need to be extremely lightweight, and thus the single hidden layer-based NN is highly desired. Therefore, we propose a fusion learning scheme to take full advantage of the multimodal amplitude information and lightweight network architecture.

To the best of our knowledge, the solution of applying the bi-directional channel reciprocity and multimodal feature-level fusion for 1-bit CS-based superimposed CSI feedback has not been investigated. The main contributions of this letter are summarized as follows:
\begin{enumerate}
  \item We exploit the multiple modals of amplitude feature from the same received signal. From the bi-directional channel reciprocity, an amplitude-learning network (AMPL-NET) is developed to capture the amplitude correlation between uplink and downlink CSI in the angle domain. In addition to the reconstructed amplitude from 1-bit CS, the multimodal amplitudes are extracted without adding additional hardware devices or reception overhead.

  \item We develop simplified reconstruction and lightweight NNs to reduce processing delay. The simplified reconstruction, lightweight AMPL-NET, and amplitude-fusion network (AMPF-NET) are employed to reduce the processing delay of original iteration reconstruction. Although both of the retrieved and reconstructed amplitudes are inaccurate, the lightweight AMPF-NET plays a critical role in the accuracy improvement. The retrieved amplitude helps to improve the reconstructed amplitudes and vice versa.

  \item We develop a paradigm of fusion learning to capture the solutions of both non-NN and NN-based receivers from a multimodal perspective for 1-bit CS-based superimposed CSI feedback. Relative to the iteration reconstruction or data-driven NN, this paradigm presents significant superiority in reducing the processing delay and computational complexity, and improving the accuracy for downlink CSI recovery, which facilitates its practical applications.

\end{enumerate}

\begin{figure}[t]
\centering
\includegraphics[width=3.5in]{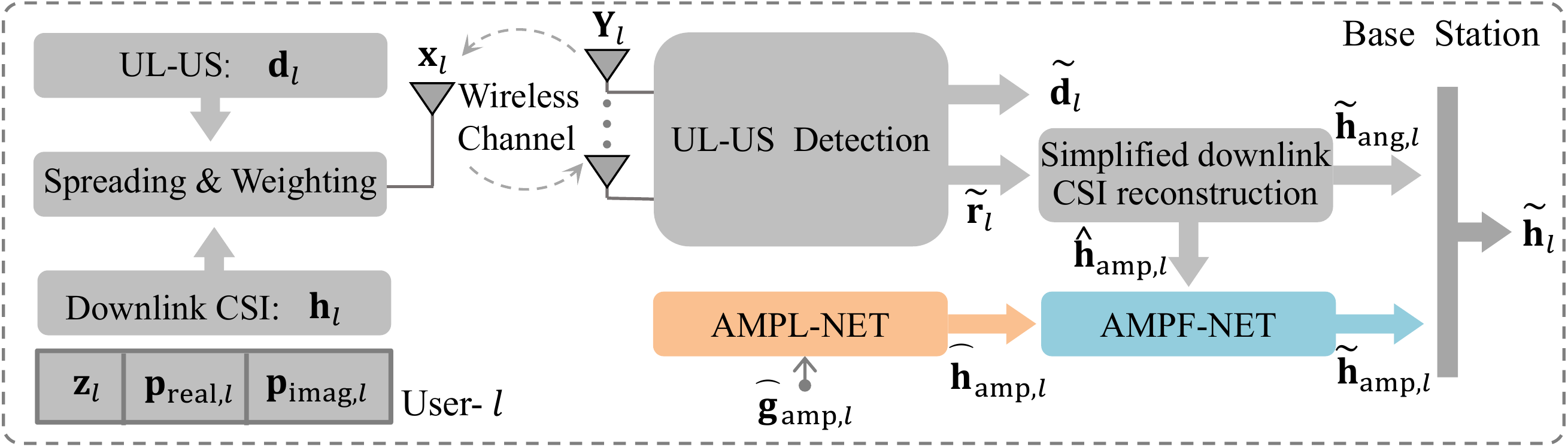}
\caption{System model.\label{fig1}}
\end{figure}

\textit{Notation}: Boldface upper case and lower case letters denote matrix and vector respectively. ${\left(\cdot \right)^T}$ denotes transpose; $\mathbf{I}_P$ is the identity matrix of size $P \times P$; ${\left\|  \cdot  \right\|}$ is the Euclidean norm; ${{\mathrm{sign}}}(\cdot)$ denotes an operator of taking symbolic information, e.g., the sign function returns 1 for positive numbers and 0 otherwise; ${{\mathrm{Re}}}(\cdot)$ and ${{\mathrm{Im}}}(\cdot)$ represent the operation of taking the real and imaginary parts of a complex value, respectively; $|\cdot|$ denotes the operation of taking the modulus of a complex value. $E[\cdot]$ represents the expectation operation. $\odot $ represents the Hadamard product. $e$ is a constant ( $e = 2.7183...$).

\section{System Model}
Considering an FDD mMIMO system that consists of a BS with $N$ antennas and $L$ single-antenna users, the system model is given in Fig.~\ref{fig1}. On the user-$l$ side, $l=1,2,\cdots,L$, the $K$-sparsity downlink CSI ${\mathbf{h}}_{l}\in \mathbb{C}^{1\times N}$ in the angular domain is compressed according to the 1-bit CS technique \cite{r1}, i.e.,
\begin{equation}
\label{equ2}
\left\{ {\begin{array}{*{20}{c}}
  {{{\mathbf{p}}_{{\textrm{real,}}l}} = {\mathrm{sign}}\left( {{\mathrm{Re}}\left( {{{\mathbf{h}}_l}{{\mathbf{\Phi }}_l}} \right)} \right)} \\
  {{{\mathbf{p}}_{{\textrm{imag,}}l}}= {\mathrm{sign}}\left( {{\mathrm{Im}}\left( {{{\mathbf{h}}_l}{{\mathbf{\Phi }}_l}} \right)} \right)}
\end{array}} \right.,
\end{equation}
with ${{\mathbf{p}}_{{\textrm{real,}}l}}$ and ${{\mathbf{p}}_{{\textrm{imag,}}l}}$ being the real and imaginary parts of the compressed downlink CSI, respectively. The $\mathbf{\Phi }_l$ denotes the $N \times M$ measurement matrix \cite{y1}. For better performance, the support-set ${\mathbf{z}}_l \in \{ 0,1 \} ^{1\times N}$ is employed to mark indexes of zero elements and non-zero elements of the downlink CSI \cite{y1}. According to $\mathbf{z}_l $, ${{\mathbf{p}}_{{\textrm{real,}}l}}$, and ${{\mathbf{p}}_{{\textrm{imag,}}l}}$, the $feedback$ $vector$ (FV) is constructed as ${{\mathbf{w}}_l} = [{{\mathbf{z}}_l},{{\mathbf{p}}_{{\textrm{real,}}l}},{{\mathbf{p}}_{{\textrm{imag,}}l}}]$, where $\mathbf{w}_l $ is a $1\times K$ vector with $K = N + M*2$. It is noteworthy that the $\mathbf{w}_l $ is a bit stream format, because the elements of $\mathbf{w}_{l}$ only contain 0 and 1. With the digital modulation, the modulated feedback vector $\mathbf{r}_{l}$ with length $T$ is obtained as
\begin{equation}
\label{equ:modulated_signal_x}
{{\bf{r}}_l} \buildrel \Delta \over = {f_{{\textrm{modu}}}}({{\bf{w}}_l}),
\end{equation}
where $f_{\textrm{modu}}(\cdot)$ denotes the mapping function of digital modulation, such as the quadrature phase shift keying (QPSK), and $T = \left\lceil {K/2} \right\rceil $ for QPSK modulation. Similar to \cite{y1}, the spreading spectrum method is employed to reduce superimposed interference. Then, the superimposition signal $\mathbf{s}_{l}$ is written as $\mathbf{s}_{l} = \frac{1}{\sqrt{T}}\mathbf{r}_{l}\mathbf{Q}_{l}^{{T}}$,
where $\mathbf{Q}_{l}\in\mathbb{R}^{P\times T}$ denotes the spreading matrix, satisfying $\mathbf{Q}_{l}^{T}\mathbf{Q}_{l}=P\mathbf{I}_T$. With the superimposition signal and the UL-US signal, the transmitted signal ${\mathbf{x}}_{l}\in \mathbb{C}^{1\times P}$ is given by \cite{y1}
\begin{equation}
\begin{aligned}
\label{equ:transmitting_signal_X}
\mathbf{x}_{l} =& \sqrt{\rho E_{l}}\mathbf{s}_{l} + \sqrt{(1-\rho)E_{l}}\mathbf{d}_{l},
\end{aligned}
\end{equation}
where $\rho \in [0,1]$ stands for the power proportional coefficient of downlink CSI, $E_{l}$ represents the
transmitted power of user-$l$, and $\mathbf{d}_l\in\mathbb{C}^{1\times P}$ denotes the modulated UL-US signal.
Without loss of generality, due to the main task of uplink services \cite{y1}, the length of UL-US is larger than that of the modulated feedback vector, i.e., $P>T$.
At the BS, after the processing of matched-filter, the received signal $\mathbf{Y}_l$ of the user-$l$, is given by
\begin{equation}
\begin{aligned}
\label{equ:received_signal_Y}
\mathbf{Y}_{l} =& \mathbf{g}_{l}\mathbf{x}_{l}  + \mathbf{N}_l,
\end{aligned}
\end{equation}
where $\mathbf{N}_l\in \mathbb{C}^{N\times P}$ represents the circularly symmetric complex Gaussian (CSCG) noise with zero-mean and variance ${\sigma_{l}^2}$ for each feedback link, ${\mathbf{x}}_{l}$ denotes the transmitted signal of the $l$-th user, and ${{\mathbf{g}}_l} = {\left[ {{g_{l,1}},{g_{l,2}}, \ldots ,{g_{l,N}}} \right]^T} \in {\mathbb{C}^{N \times 1}}$ denotes the uplink channel matrix from the user-$l$ to the BS.
With the received signal $\mathbf{Y}_{l}$, the UL-US detection is utilized to detect the UL-US ${\bf{\widetilde d}}_{l}$ and modulated feedback vector ${\bf{\widetilde r}}_{l}$. Then, the simplified downlink CSI reconstruction is employed to extract the initial amplitude feature ${\bf{\widehat h}}_{{\textrm{amp}},l}$ and angle feature ${\bf{\widetilde h}}_{{\textrm{ang}},l}$. Meanwhile, the AMPL-NET is designed to learn the downlink CSI amplitude feature according to bi-directional channel reciprocity. Subsequently, we develop the AMPF-NET to fuse the amplitude features based on the simplified downlink CSI reconstruction and AMPL-NET. Finally, the amplitude fused from AMPF-NET is combined with its corresponding ${\bf{\widetilde h}}_{{\textrm{ang}},l}$ as a fully recovered downlink CSI ${{\bf{\widetilde h}}_l}$.

\begin{figure}[t]
\centering
\includegraphics[width=3.5in]{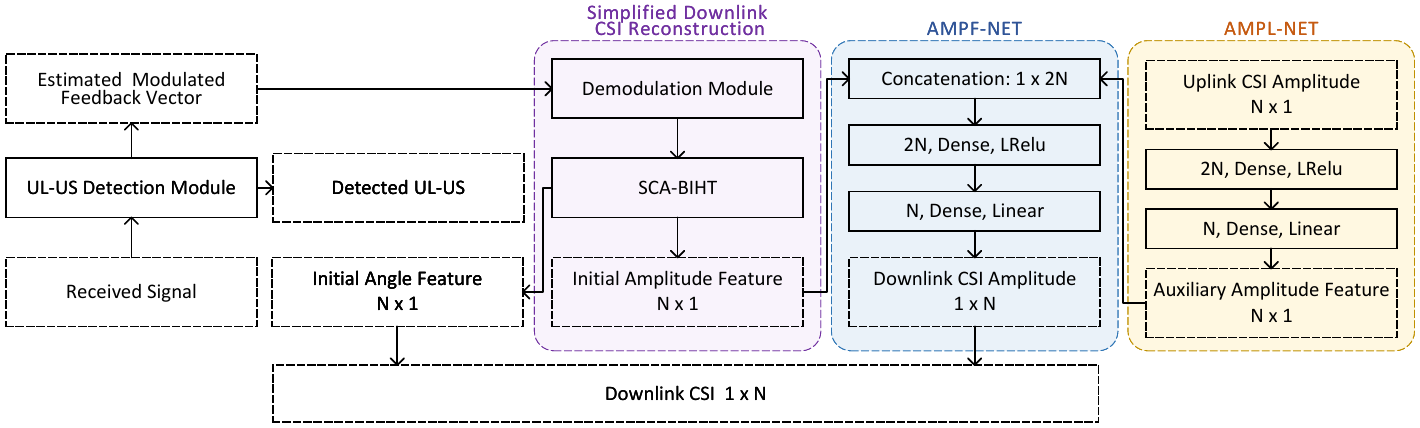}
\caption{The fusion learning-based feedback scheme.}\label{figlc}
\end{figure}

\section{Fusion Learning-based Feedback Scheme}
\label{sec:guidelines}

The fusion learning-based feedback scheme is shown in Fig.~\ref{figlc}, in which a simplified reconstruction method is first employed to reconstruct the downlink CSI, which is presented in Section III-A.
Then, we develop AMPL-NET and AMPF-NET to learn and fuse the downlink CSI amplitude, respectively. AMPL-NET learns the downlink CSI amplitude from uplink CSI, and AMPF-NET follows the idea of multimodal feature-level fusion. Both AMPL-NET and AMPF-NET are elaborated in Section III-B. In Section III-C, the analysis of computational complexity is given.

\subsection{Simplified Downlink CSI Reconstruction}
After the UL-US detection, the FV $\widetilde{\mathbf{w}}_l$ is recovered from the detected modulated feedback vector ${\bf{\widetilde r}}_{l}$ given in \eqref{equ:modulated_signal_x}.
We employ the reconstruction algorithm with $\widetilde{\mathbf{w}}_l$ and $\alpha$ iterations to perform a simplified version of the downlink CSI reconstruction, i.e.,
\begin{equation}
\begin{aligned}
\label{equ:received_signal_Y}
\left[ {\begin{array}{*{20}{c}}
  {\bf{\widehat h}}^{T}_{{\textrm{amp}},l} , {\bf{\widetilde h}}^{T}_{{\textrm{ang}},l}
\end{array}} \right]^{T}=f_{\textrm{SCA-BIHT}}(\widetilde{\mathbf{w}}_l, \alpha),
\end{aligned}
\end{equation}
where $f_{\textrm{SCA-BIHT}}(\cdot)$ denotes the SCA-BIHT reconstruction algorithm \cite{y1}, and $\alpha \leq 5$ is considered in our work, which is far less than the iteration times of the iterative method in \cite{y1}. Then, the features of downlink CSI are extracted, i.e., the initial amplitude feature ${\bf{\widehat h}}_{{\textrm{amp}},l}$ and angle feature ${\bf{\widetilde h}}_{{\textrm{ang}},l}$ of the downlink CSI are obtained for subsequent recovery.

\subsection{AMPL-NET and AMPF-NET}

In order to obtain the downlink CSI amplitude feature which is different from the conventional feedback perspective, we construct the lightweight and effective AMPL-NET, which utilizes the bi-directional correlation of CSI amplitude\cite{d3}.
Then, a certain downlink CSI amplitude feature, which is called auxiliary amplitude feature ${{\bf{\mathord{\buildrel{\lower3pt\hbox{$\scriptscriptstyle\frown$}}
\over h} }}_{{\textrm{amp}},l}}$, is learned from AMPL-NET to complement the initial amplitude feature ${\bf{\widehat h}}_{{\textrm{amp}},l}$. Next, to refine the amplitude of the downlink CSI, we borrow the idea of multimodal feature-level fusion and design AMPF-NET, which fuses initial amplitude feature (from the simplified reconstruction method in \cite{y1}) and auxiliary amplitude feature (from AMPL-NET).

\subsubsection{Network Design}
According to \cite{a17}, choosing the number of hidden neurons or layers is still a challenge in the NN. Based on a large number of experiments, we design the AMPL-NET and AMPF-NET, both of which are composed of one input layer, one hidden layer, and one output layer.  The network architectures of AMPL-NET and AMPF-NET are summarized in Table~\ref{table_I}, and the detailed descriptions are given as follows.

In AMPL-NET, neuron numbers of the input layer, hidden layer, and output layer are $N$, $2N$, and $N$, respectively, while in AMPF-NET they are $2N$, $2N$, and $N$, respectively.
A batch normalization is employed to normalize the input sets in both AMPL-NET and AMPF-NET, forming the network input as zero mean and unit variance, and the leaky rectified linear unit (LReLU) \cite{a15} and linear activation are employed as activation functions for the hidden layer and output layer, respectively.
For training the AMPL-NET, the uplink channel ${{\mathbf{g}}_l}$ in time domain is transformed to the angular domain\cite{cc1}. By denoting the transformed uplink channel as ${{\mathbf{\overset{\lower0.5em\hbox{$\smash{\scriptscriptstyle\frown}$}}{g} }}_l}$, its amplitude is given by
\begin{equation}\label{EQ21}
{{\mathbf{\overset{\lower0.5em\hbox{$\smash{\scriptscriptstyle\frown}$}}{g} }}_{{\textrm{amp}},l}} = {\left[ {\left| {{{\overset{\lower0.5em\hbox{$\smash{\scriptscriptstyle\frown}$}}{g} }_{l,1}}} \right|,\left| {{{\overset{\lower0.5em\hbox{$\smash{\scriptscriptstyle\frown}$}}{g} }_{l,2}}} \right|, \ldots ,\left| {{{\overset{\lower0.5em\hbox{$\smash{\scriptscriptstyle\frown}$}}{g} }_{l,N}}} \right|} \right]^T}.
\end{equation}
Using ${{\mathbf{\overset{\lower0.5em\hbox{$\smash{\scriptscriptstyle\frown}$}}{g} }}_{{\textrm{amp}},l}}$ as the input of the AMPL-NET, the auxiliary feature ${{\bf{\mathord{\buildrel{\lower3pt\hbox{$\scriptscriptstyle\frown$}}
\over h} }}_{{\textrm{amp}},l}}$ of the downlink CSI amplitude is obtained according to
\begin{equation}\label{EQ22}
{{\bf{\mathord{\buildrel{\lower3pt\hbox{$\scriptscriptstyle\frown$}}
\over h} }}_{{\textrm{amp}},l}} = f_{\textrm{AMPL}}({{\bf{\mathord{\buildrel{\lower3pt\hbox{$\scriptscriptstyle\frown$}}
\over g} }}_{{\textrm{amp}},l}}, \Theta_{\textrm{AMPL}}),
\end{equation}
where $f_{\textrm{AMPL}}(\cdot)$ and $\Theta_{\textrm{AMPL}}$ denote the amplitude learning operation and its network parameter, respectively.

In AMPF-NET, the input ${{\mathbf{\overset{\lower0.5em\hbox{$\smash{\scriptscriptstyle\smile}$}}{h} }}_{{\textrm{amp}},l}} \in {\mathbb{R}^{1 \times 2N}}$ is spliced by ${\bf{\widehat h}}_{{\textrm{amp}},l}$ and ${{\bf{\mathord{\buildrel{\lower3pt\hbox{$\scriptscriptstyle\frown$}}
\over h} }}_{{\textrm{amp}},l}}$ , i.e.,
\begin{equation}\label{EQ24}
{{\mathbf{\overset{\lower0.5em\hbox{$\smash{\scriptscriptstyle\smile}$}}{h} }}_{{\textrm{amp}},l}} = [{\bf{\widehat h}}_{{\textrm{amp}},l}^T,{{\bf{\mathord{\buildrel{\lower3pt\hbox{$\scriptscriptstyle\frown$}}
\over h} }}^T_{{\textrm{amp}},l}}].
\end{equation}
Then, using the AMPF-NET, the amplitude of the downlink CSI ${\bf{\widetilde h}}_{{\textrm{amp}},l}$ is obtained by
\begin{equation}\label{EQ25}
{\bf{\widetilde h}}_{{\textrm{amp}},l} = f_{\textrm{AMPF}}({{\mathbf{\overset{\lower0.5em\hbox{$\smash{\scriptscriptstyle\smile}$}}{h} }}_{{\textrm{amp}},l}}, \Theta_{\textrm{AMPF}}),
\end{equation}
where $f_{\textrm{AMPF}}(\cdot)$ and $\Theta_{\textrm{AMPF}}$ denote the amplitude fusion operation and its network parameter, respectively.

\subsubsection{Training and Deployment}
The training sets are acquired by simulation, and a significant amount of data samples are collected to train AMPL-NET and AMPF-NET, respectively. Specifically, these data samples are generated as follows.

\begin{table}[]

\renewcommand{\arraystretch}{1.2}
\caption{Architecture Of AMPL-NET and AMPF-NET}
\label{table_I}
\centering
\scalebox{0.66}{
\begin{tabu}{@{}c|c|c|c|c|c|c@{}}
\tabucline[0.8pt]{-}
\multirow{2}{*}{Layer} & \multicolumn{2}{c|}{Input} & \multicolumn{2}{c|}{Hidden} & \multicolumn{2}{c}{Output} \\ \cline{2-7}
 & \multicolumn{1}{c|}{AMPL-NET} & \multicolumn{1}{c|}{AMPF-NET} & \multicolumn{1}{c|}{AMPL-NET} & \multicolumn{1}{c|}{AMPF-NET} & \multicolumn{1}{c|}{AMPL-NET} & \multicolumn{1}{c}{AMPF-NET} \\ \tabucline[0.8pt]{-}
Batch normalization &$\surd$ &$\surd$ & $\times$ & $\times$ &$\times$ &$\times$ \\ \hline
Neuron number       & $N$    & $2N$     & $2N$  & $2N$    & $N$     & $N$    \\ \hline
Activation function & None & None & LReLU & LReLU & Linear & Linear  \\ \tabucline[0.8pt]{-}
\end{tabu}}
\end{table}

The bi-directional channels are generated by MATLAB 5G Toolbox, which is subject to specifications of the Clustered-Delay-Line (CDL) channel model in 3GPP TR 38.901 \cite{tr1}. Similar to the setting in \cite{cc1}, the frequency-independent parameters (e.g., the azimuth angle of departure (AoD)) are fixed, while varying the complex gain of each path between the downlink and uplink channels.
Then, ${{\mathbf{\overset{\lower0.5em\hbox{$\smash{\scriptscriptstyle\frown}$}}{g} }}_l}$ and ${{\mathbf{h}}_l}$ are obtained by transforming the generated uplink and downlink channels to the angular domain, respectively\cite{cc1}.
To train the AMPL-NET, we form input sets according to (\ref{EQ21}), while we collect amplitude features ${\bf{\widehat h}}_{{\textrm{amp}},l}$ and ${{\bf{\mathord{\buildrel{\lower3pt\hbox{$\scriptscriptstyle\frown$}}
\over h} }}_{{\textrm{amp}},l}}$ of the downlink CSI to form input sets according to (\ref{EQ24}) to train the AMPF-NET. Then, we save the corresponding ${\bf{h}}_{{\textrm{amp}},l}$ from ${{\mathbf{h}}_l}$ as target sets of AMPL-NET and AMPF-NET, respectively.
The optimization goal of AMPL-NET is to minimize the mean squared error (MSE) between ${{\bf{\mathord{\buildrel{\lower3pt\hbox{$\scriptscriptstyle\frown$}}
\over h} }}_{{\textrm{amp}},l}}$ and ${{\mathbf{h}}_{{\text{amp}},l}}$, which is derived as
\begin{equation}\label{EQ20}
\begin{gathered}
\mathop {\min }\limits_{{\Theta _{{\text{AMPL}}}}} E\left[ {{{\left\| {f_{\textrm{AMPL}}({{\bf{\mathord{\buildrel{\lower3pt\hbox{$\scriptscriptstyle\frown$}}
\over g} }}_{{\textrm{amp}},l}}, \Theta_{\textrm{AMPL}}) - {{\mathbf{h}}_{{\text{amp}},l}}} \right\|}^2}} \right].
\end{gathered}
\end{equation}
Similarly, the AMPF-NET minimizes the MSE of the fused amplitude, i.e., $E\left[ {{{\left\| {{{\bf{\widetilde h}}_{{\textrm{amp}},l} - {{\mathbf{h}}_{{\text{amp}},l}}}} \right\|}^2}} \right]$, which is further expressed by
\begin{equation}\label{EQ20}
\begin{gathered}
\mathop {\min }\limits_{{\Theta _{{\text{AMPF}}}}} E\left[ {{{\left\| {f_{\textrm{AMPF}}({{\mathbf{\overset{\lower0.5em\hbox{$\smash{\scriptscriptstyle\smile}$}}{h} }}_{{\textrm{amp}},l}}, \Theta_{\textrm{AMPF}}) - {{\mathbf{h}}_{{\text{amp}},l}}} \right\|}^2}} \right].
\end{gathered}
\end{equation}
We perform the training once for both AMPL-NET and AMPF-NET, and save the trained network parameters for testing.

By using the AMPF-NET, the high precision downlink amplitude ${\bf{\widetilde h}}_{{\textrm{amp}},l}$ is obtained.
Then, the amplitude ${\bf{\widetilde h}}_{{\textrm{amp}},l}$ is combined with its corresponding angle ${\bf{\widetilde h}}_{{\textrm{ang}},l}$ from the simplified downlink CSI reconstruction, and this leads to the recovered downlink CSI ${{\bf{\widetilde h}}_l}$, i.e., ${{\bf{\widetilde h}}_l} = {\bf{\widetilde h}}_{{\textrm{amp}},l} \odot {e^{j{\bf{\widetilde h}}^T_{{\textrm{ang}},l}}}$.
Compared with the 1-bit CS-based superimposed CSI feedback scheme in \cite{y1}, the proposed scheme demonstrates a better CSI reconstruction accuracy, and reduces the online running time and computational complexity by balancing the off-line training and online running.

\subsection{Complexity Analysis}
For description convenience, ``Proposed'' is used to denote the proposed scheme; ``Ref\cite{y1}'' represents the conventional 1-bit superimposed feedback in \cite{y1}; ``Ref\cite{r1}'' denotes the 1-bit CS feedback method with TDM mode in \cite{r1}. The comparison of computational complexity is given in Table~\ref{table_III}, where $\beta$ denotes the iteration times required for the downlink CSI reconstruction schemes in \cite{y1} and \cite{r1}. SCA-BIHT has the computational complexity of $4MN$ for each iteration \cite{y1}. The proposed simplified version of the downlink CSI reconstruction has fewer iterations than \cite{y1} and \cite{r1}, i.e., $\alpha < \beta$. In addition, weight number and floating-point operations (FLOPs) are the most common metrics to describe the NN complexity \cite{cc1}. From \cite{cc1}, the total NN weight number of the proposed AMPL-NET and AMPF-NET is $10N^{2}+6N$, and the total FLOPs number is $20N^{2}-6N$. Thus, the total complexity of the proposed scheme (including simplified downlink CSI reconstruction, AMPL-NET, and AMPF-NET) is $4MN\cdot{\alpha}+10N^{2}+6N+20N^{2}-6N=4MN\cdot{\alpha}+30N^{2}$.
For the case where $M = 2N$, $\alpha = 5$, and $\beta = 100$, the computational complexities of the downlink CSI reconstruction in ``Proposed'', ``Ref\cite{y1}'', and ``Ref\cite{r1}'' are $70N^{2}$, $800N^{2}$, and $800N^{2}+400N$, respectively. Therefore, the proposed scheme has lower computational complexity than those of \cite{y1} and \cite{r1}.

\begin{table}[]
\renewcommand{\arraystretch}{1.2}
\caption{Computational Complexity Of Downlink CSI Reconstruction}
\label{table_III}
\centering
\scalebox{0.72}{
\begin{tabu}{@{}c|c|c|c@{}}
\tabucline[0.55pt]{-}
Method       & Proposed  &  Ref\cite{y1}  &  Ref\cite{r1}                           \\ \Xhline{0.8pt}
Complexity      & $4MN\cdot{\alpha}+30N^{2}$  &$4MN\cdot{\beta}$  &$4(M+1)N\cdot{\beta}$   \\ \hline
Case ($M = 2N$)  & $70N^{2}$ ($\alpha = 5$)  & $800N^{2}$ ($\beta = 100$) & $800N^{2}+400N$ ($\beta = 100$)  \\ \tabucline[0.55pt]{-}
\end{tabu}}

\end{table}

\section{Experiment results}
In this section, we provide numerical results of the proposed scheme. Definitions and basic parameters involved in simulations are first given in Section IV-A. Then, to verify the effectiveness of the proposed scheme, the normalized mean squared error (NMSE) of the reconstructed downlink CSI and the online running time are given in Section IV-B. Finally, the robustness of the proposed scheme is verified in Section IV-C. It should be noted that, the conventional superimposition scheme in \cite{y1} needs to verify the bit error rate (BER) of UL-US. Considering that the same BER can be obtained as that of \cite{y1} due to the same detection scheme, we here only verify the NMSE for the recovery of downlink CSI.

\subsection{Parameters Setting}
Definitions involved in simulations are given as follows. The equivalent signal-to-noise ratio (SNR) ratio and NMSE are defined the same way as \cite{y1}.
During the experiments, $P=512$, $N=64$, and the compression rate $c$ is defined as $c=M/N$.
We adopt the Walsh matrix as the spreading matrix ${{\mathbf{Q}}_l}$ \cite{y1}. The UL-US ${\mathbf{d}}_l$ is formed by using QPSK modulation.
For AMPL-NET and AMPF-NET, training data-sets are generated by (\ref{EQ21}) and (\ref{EQ24}) respectively, and testing data-sets are generated the same as training data-sets.
AMPL-NET and AMPF-NET are trained under the noise-free setting.
Training, validation, and testing sets have $70,000$, $15,000$, and $15,000$ samples, respectively.
In this letter, the NMSE performance of the proposed scheme is compared with those of \cite{y1} and \cite{r1}.

\begin{figure}[t]
\centering
\includegraphics[width=3.5in]{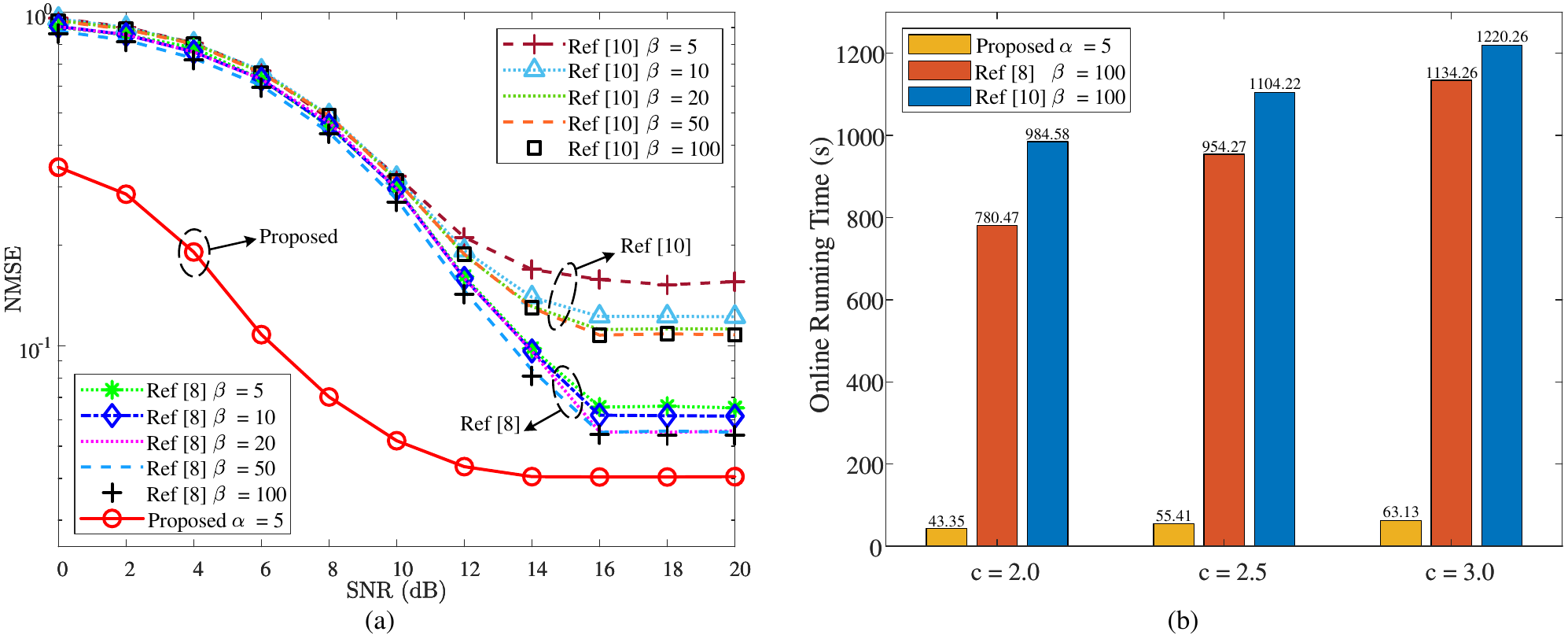}
\caption{(a) NMSE versus SNR, where $P{\rm{ = 512}}$, $c = 2.0$, and $\rho = 0.10$. (b) Online running time comparison of ``Proposed'', ``Ref\cite{y1}'', and ``Ref\cite{r1}''for $10^5$ experiments, where $P{\rm{ = 512}}$, $\rho = 0.20$, $\alpha = 5$, and $\beta = 100$ is in both ``Ref\cite{y1}'' and ``Ref\cite{r1}''.}\label{fig2}
\end{figure}

\subsection{NMSE Performance and Online Running Time}
\subsubsection{NMSE Performance}

To validate the effectiveness of the proposed reconstruction scheme, NMSE curves of the downlink CSI recovered by ``Proposed'', ``Ref\cite{y1}'', and ``Ref\cite{r1}'' are given in Fig.~\ref{fig2} (a).
In Fig.~\ref{fig2} (a), the NMSE curves in terms of the downlink CSI's recovery are presented, where $c=2.0$ and $\rho=0.10$ are considered.
The ``Proposed'' employs $5$ iterations of the simplified downlink CSI reconstruction for feature extraction, i.e., $\alpha = 5$.
By contrast, different iteration times (i.e., $\beta = 5$, $\beta = 10$, $\beta = 20$, $\beta = 50$, and $\beta = 100$) are given for downlink CSI reconstructions of ``Ref\cite{y1}'' and ``Ref\cite{r1}''.
From Fig.~\ref{fig2} (a), the downlink CSI's NMSE of ``Proposed'' is smaller than those of the ``Ref\cite{y1}'' and ``Ref\cite{r1}'' in the whole SNR regions. For example, when ${\textrm{SNR}}=10$dB, the NMSE of ``Proposed'' is about $5.2\times 10^{-2}$, while ``Ref\cite{y1}'' and ``Ref\cite{r1}'' with $\beta = 5$ reach about $2.9\times 10^{-1}$ and $3.2\times 10^{-1}$, respectively.
In a word, the proposed scheme shows a better NMSE performance in all given SNR regions.
\subsubsection{Online Running Time}

To illustrate the improvement of processing time, the online running time of ``Proposed'', ``Ref\cite{y1}'', and ``Ref\cite{r1}'' is compared in Fig.~\ref{fig2} (b) with $c$ varying from $2.0$ to $3.0$. Especially, the ``Ref\cite{y1}'' and ``Ref\cite{r1}'' follow the setting of $\beta=100$ in Fig.~\ref{fig2} (a). For a fair comparison, $10^5$ times of online running experiments are separately conducted for ``Proposed'', ``Ref\cite{y1}'', and ``Ref\cite{r1}'' on the same personal computer (with CPU i5-8250U) by using MATLAB software. For each given $c$, the online running time of ``Proposed'' is shorter than those of ``Ref\cite{y1}'' and ``Ref\cite{r1}''. For instance, when $c=3.0$, the online running time of ``Proposed'', ``Ref\cite{y1}'', and ``Ref\cite{r1}'' are $63.13s$, $1134.26s$, and $1220.26s$, respectively. This reflects that the proposed scheme reduces online running time. The reason for the improvement is that the trade-off is achieved by balancing the off-line training and online running.
In addition, as $c$ varies from $2.0$ to $3.0$, the online running time of ``Proposed'', ``Ref\cite{y1}'', and ``Ref\cite{r1}'' arises due to the increase of computational complexity.
On the whole, with the low running time and computational complexity in Table~\ref{table_III}, the proposed scheme effectively reduces the processing delay compared with ``Ref\cite{y1}'' and ``Ref\cite{r1}''.

\subsection{Robustness Evaluation}
\begin{figure}[t]
\centering
\includegraphics[width=3.5in]{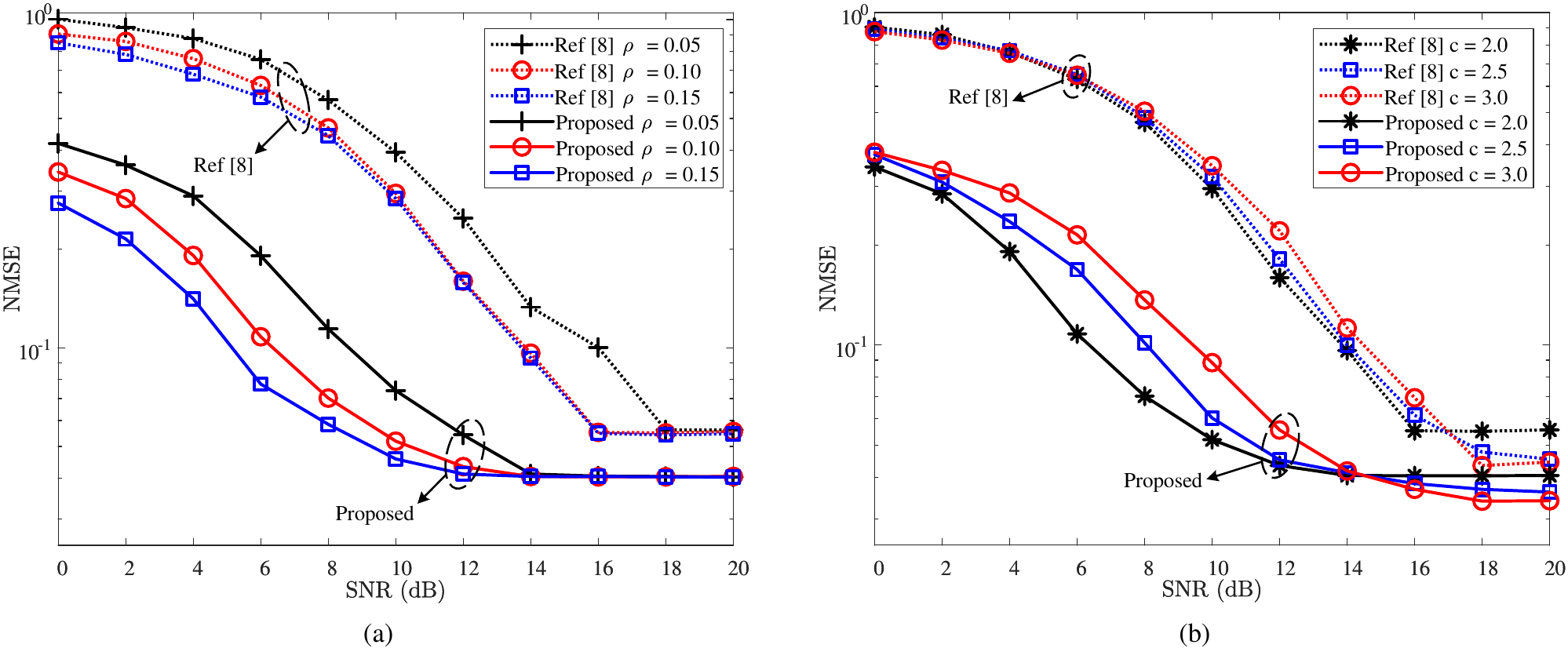}
\caption{NMSE versus SNR, where $P{\rm{ = 512}}$, $\alpha = 5$, and $\beta = 20$, (a) $c = 2.0$, (b) $\rho = 0.10$.}\label{fig3}
\end{figure}

To verify the robustness of NMSE performance against the impact of $\rho$, the NMSE curves with variant $\rho$ (i.e., $\rho=0.05$, $\rho=0.10$, and $\rho=0.15$) are plotted in Fig.~\ref{fig3} (a). For each given $\rho$, the downlink CSI's NMSE of the ``Proposed'' is smaller than that of the ``Ref\cite{y1}''. As the increase of $\rho$ (increases from $0.05$ to $0.15$), the NMSEs decrease for both ``Ref\cite{y1}'' and ``Proposed'', and vice versa. The reason is that the downlink CSI could obtain more transmission power with a larger value of $\rho$. In addition, with the increase of SNR, the curves gradually converge for the reason that the main influence of NMSE comes from the superimposed interference in the relatively high SNR region. On the whole, for each given value of $\rho$, the NMSE of ``Ref\cite{y1}'' is reduced by the ``Proposed'' in all given SNR regions. Thus, the proposed scheme shows its robustness in improving the NMSE performance against the impact of $\rho$.

To further validate the robustness against the impact of $c$, Fig.~\ref{fig3} (b) depicts the NMSE performance of the downlink CSI recovery with compression rate $c$ varying from $2.0$ to $3.0$.
In Fig.~\ref{fig3} (b), for each given $c$, the NMSE of the ``Proposed'' is smaller than that of the ``Ref\cite{y1}''.
Furthermore, when ${\textrm{SNR}}\leq14$dB, the NMSE of ``Proposed'' increases as the increase of $c$ (i.e., $c$ increases from $2.0$ to $3.0$). The possible reason is that the higher compression rate results in lower spreading gain (i.e., $P/M$).
In the low SNR region, the NMSE performance is mainly impacted by noise interference and limited by the low spread spectrum gain.
However, the NMSE's convergence value of high compression rate is smaller than that of low compression rate.
For example, in the case of $c=2.0$, $c=2.5$, and $c=3.0$, the NMSE's convergence values of ``Proposed'' are about $4.03 \times 10^{-2}$, $3.61 \times 10^{-2}$, and $3.39 \times 10^{-2}$, respectively. The reason for the analysis is that the higher compression rate brings more reconstruction information in the high SNR region, where the noise interference almost disappeared.
On the whole, for each given value of $c$ , the NMSE of ``Ref\cite{y1}'' is reduced by the ``Proposed''.
Thus, the proposed scheme possesses its robustness against the impact of $c$.

To sum up, from Fig.~\ref{fig2} (a), the downlink CSI's NMSE performance of ``Ref\cite{y1}'' and ``Ref\cite{r1}'' is effectively improved by the ``Proposed''. Moreover, Fig.~\ref{fig2} (b) demonstrates that the proposed scheme has a shorter online running time, thereby obtaining a lower processing delay. In addition, Fig.~\ref{fig3} (a) and Fig.~\ref{fig3} (b) show that the improvement of ``Proposed'' are robust against the impacts of $\rho$ and $c$, respectively.

\section{Conclusion}
This letter presents a fusion learning scheme for 1-bit CS-based superimposed CSI feedback for mMIMO wireless communications.
With a simplified downlink CSI reconstruction and a lightweight AMPL-NET, the initial and auxiliary features of the downlink CSI amplitude are extracted and then refined by a multimodal feature-level AMPF-NET. Experiments show that, compared with the conventional 1-bit CS-based superimposed CSI feedback, the proposed scheme achieves a significant improvement on NMSE performance of the downlink CSI recovery. Besides, it presents a reduction in processing delays and robustness against parameter variations.

\appendices

\ifCLASSOPTIONcaptionsoff
  \newpage
\fi

\end{document}